\documentclass[10pt]{article}
\usepackage{graphicx}
\usepackage{floatflt}
\usepackage{subfigure}
\usepackage{array}
\usepackage{indentfirst}
\begin{document}
\newcommand{\eqnzero}{\setcounter{equation}{0}} 
\newcommand{\bq}{\begin{equation}}
\newcommand{\eq}{\end{equation}}
\newcommand{\bqa}{\begin{eqnarray}}
\newcommand{\eqa}{\end{eqnarray}}
\newcommand{\nll}{\nonumber\\}
\newcommand{\Litwo}{\mbox{${\rm{Li}}_{2}$}}
\newcommand{\ds }{\displaystyle}
\newcommand{\sss}[1]{\scriptscriptstyle{#1}}
\newcommand{\eqn}[1]{Eq.~(\ref{#1})}
\newcommand{\lk}{\hspace{-3mm}}
\vspace*{40mm}
\begin{center}
{\LARGE\bf Standard Model light-by-light scattering in {\tt SANC}: analytic and numeric evaluation.}
\vspace*{15mm}

{\bf D.~Bardin, L.~Kalinovskaya, E.~Uglov}
\vspace*{10mm}
{\normalsize{\it

        Dzhelepov Laboratory for Nuclear Problems, JINR,   \\
        ul. Joliot-Curie 6, RU-141980 Dubna, Russia       }}
\vspace*{20mm}
\end{center}
\begin{abstract}
\noindent
In this paper we describe the implementation of the SM process $\gamma\gamma\to\gamma\gamma$ through a fermion and boson loops into the framework of {\tt SANC} system. The computations of this process takes into account non-zero mass of loop particles. We briefly describe additional precomputation modules used for calculation of massive box diagrams. We present the covariant and helicity amplitudes for this process, some particular cases of $D_0$ and $C_0$ Passarino--Veltman functions and also numerical results of corresponding {\tt SANC} module evaluation. Whenever possible, we compare the results with those existing in the literature.
\end{abstract}
\vspace*{50mm}
\footnoterule
{\footnotesize
E-mails: bardin@nusun.jinr.ru, kalinov@nusun.jinr.ru, corner@nusun.jinr.ru}
\clearpage

\section{Introduction \label{introduction}}

{\tt SANC} is a computer system for semi-automatic calculations of realistic and pseudo-observables for various processes of elementary particle interactions "from the SM Lagrangian to event distributions" at the one-loop precision level for the present and future colliders --- TEVATRON, LHC, electron Linear Colliders (ILC, CLIC), muon factories and others. To learn more about available processes in {\tt SANC} see the description in~\cite{Andonov:2006,Modules:2009} and look at our home pages at JINR and CERN~\cite{SANC:2006}.

Light-by-light scattering is one of the most fundamental processes. It proceeds via one-loop box diagrams containing charged particles. The first results for the QED low energy limit of this process were obtained by Euler~\cite{Euler:1936}. Then Karplus and Neumann~\cite{Karplus:1951} found a solution for QED in general but complicated way. The QED cross sections in the high energy limit were calculated by Ahiezer~\cite{Ahiezer:1981}. Nowadays there are computations for $\gamma\gamma\to\gamma\gamma$ process in the electorweak Standard Model~\cite{Jikia:1993,Jikia:1997,Bohm:1994sf,Diakonidis:2006} and even for two-loop QCD and QED corrections~\cite{Bern:2001}.

In this paper we describe the implementation of the SM process $\gamma\gamma\to\gamma\gamma$ through fermion and boson loops and corresponding precomputation modules into the framework of {\tt SANC} system. The computations of this process take into account non-zero mass of the loop particles.

One should emphasize that the obtained building blocks and procedures of precomputation for box diagrams in QED and EW (as in $\gamma\gamma\to\gamma\gamma$) is the first step in the creation of environment for implementation of the similar four-bosons processes in the Standard Model (like $\gamma\gamma\to ZH$, $\gamma\gamma\to ZZ$~\cite{Diakonidis:2006}) and in QCD (like $gg\to \gamma\gamma$, $gg\to ZZ$, $gg\to W^{+}W^{-}$ etc.).

The paper is organized as follows:

First we discuss some notations and common expression for cross section in section~\ref{XS}. 

In section~\ref{CA} we discuss diagrams for $\gamma\gamma\to\gamma\gamma$ process and covariant amplitude tensor structure. 

The helicities amplitudes approach~\cite{Andonov:2006,VW:1996} and their expressions for light-by-light scattering in general (massive) and in limiting (massless) cases are listed in section~\ref{HA}. 

In section \ref{tree} we shortly describe precomputation strategy and the place of this process on the {\tt SANC} process tree.

The implementation of analitycal results and the {\tt SANC} modules concept we describe in section \ref{module}.

At last in section \ref{results} one can find the numerical result and comparisons with those existing in the literature. 

Additionally, in Appendix section we list answers for particular cases of $D_0$, $C_0$ and $B_0$ Passarino--Veltman (PV) functions~\cite{Passarino:1979} (see also the book~\cite{Bardin:1999}), which are needed for calculation of light-by-light scattering through massive and massless loop particles. Finally, we present strings and basis for covariant amplitude.

\section{Light-by-light scattering process \label{process}}
\subsection{Notation, cross section\label{XS}}

The 4-momenta of incoming photons are denoted by $p_1$ and $p_2$, of the outgoing ones --- by $p_3$ and $p_4$. The amplitudes are calculated for scattering of real photons, that is $p_1^2=0\,,~p_2^2=0\,,~p_3^2=0\,,~p_4^2=0$. The 4-momentum conservation law reads: 
\bqa
p_1+p_2-p_3-p_4=0\,.
\eqa

The Mandelstam variables are:\footnote{Note, that in {\tt SANC} we use Pauli metric.}
\bqa
&&s=-(p_1+p_2)^2=-2 p_1 \cdot p_2\,,\qquad t=-(p_1-p_3)^2=2 p_1 \cdot p_3\,,
\nll
&&u=-(p_1-p_4)^2=2 p_1  \cdot p_4\,,\qquad\qquad s+t+u=0\,.
\eqa

For the $2\to 2$ $\gamma\gamma\to\gamma\gamma$ process, the cross section has the form:
\bqa
d\sigma_{\gamma\gamma\to\gamma\gamma}=
\frac{1}{j}\left|{\cal{A}}_{\gamma\gamma\to\gamma\gamma}\right|^2 d\Phi^{(2)}\,,
\eqa
where $j=4\sqrt{(p_1p_2)^2}$ is the flux, ${\cal{A}}_{\gamma\gamma\to\gamma\gamma}$ is the covariant amplitude (CA) of the process, and $d\Phi^{(2)}$ is the two body phase space:
\bqa
d\Phi^{(2)}=(2\pi)^4\delta\left(p_1+p_2-p_3-p_4\right) 
\frac{d^4 p_3 \delta\left( p_3^2 \right)}{(2\pi)^3}\frac{d^4p_4\delta\left( p_4^2 \right)}{(2\pi)^3}\,.
\eqa

For the differential cross section one gets: 
\bqa
d\sigma_{\gamma\gamma\to\gamma\gamma}=
\frac{1}{128\pi s}\left|{\cal{A}}_{\gamma\gamma\to\gamma\gamma}\right|^2 d \cos\theta\,,
\eqa
where $s=4\omega^2$, $\omega$ is the photons energy and $\theta$ --- the scattering angle in the center of mass system (cms). 
\subsection{Covariant amplitude \label{CA}}

The covariant one-loop amplitude corresponds to a result of the straightforward standard calculation of all diagrams contributing to a given process at Born (tree) and one-loop (1-loop) levels.
 
The CA is being represented in a certain basis, made of strings of Dirac matrices and/or external momenta (structures), contracted with polarization vectors of vector bosons, $\epsilon(p)$, if any. The amplitude also contains kinematical factors and coupling constants and is parameterized by a certain number of Form Factors (FFs), which are denoted by ${\cal F}_{i}$, in general with an index $i$ labeling the corresponding structure. The number of ${\cal F}_{i}$ is equal to the number of independent structures.

\begin{figure}[htp]
\subfigure[st-channel]{\label{fig:edge-a}\includegraphics[scale=0.45]{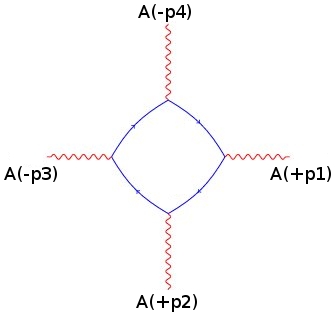}}
\subfigure[su-channel]{\label{fig:edge-b}\includegraphics[scale=0.45]{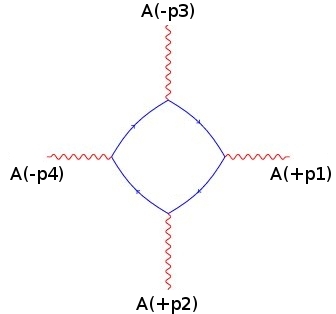}}
\subfigure[ut-channel]{\label{fig:edge-c}\includegraphics[scale=0.45]{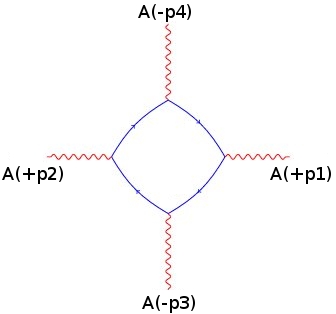}}
\caption[$\gamma\gamma\to\gamma\gamma$ process QED diagrams]
        {$\gamma\gamma\to\gamma\gamma$ process QED diagrams\label{QED.diagrams}}
\end{figure}

The $\gamma\gamma\to\gamma\gamma$ process in QFT appears due to non-linear effects of interaction with vacuum, so this process has no Born or tree level. Corresponding diagrams start from the one-loop level and in QED there are box diagrams with four internal fermions of equal mass. The number of not identical diagrams (or topologies) is equal to six. But three of them differ from another only by the orientation of the internal fermionic loop, giving the same contribution or a factor 2 to the amplitude. So, only three topologies (st, su and ut channels) remain which are related by simple permutations of external photons in the diagrams shown in Figure~\ref{QED.diagrams}: su-channel is obtained from st-channel by $p_3 \leftrightarrow p_4$ rotation and ut-channel --- by $p_2 \leftrightarrow p_3$. The sum of these fermionic diagrams is a gauge invariant in each generation of particles.

\begin{figure}[htp]
\begin{tabular}{m{5cm}m{5cm}m{5cm}}
\subfigure[box topology]{\label{fig:edge-a}\includegraphics[scale=0.45]{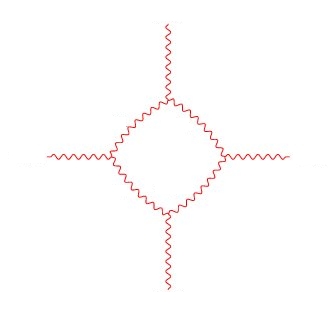}}&
\subfigure[pinch topology]{\label{fig:edge-b}\includegraphics[scale=0.45]{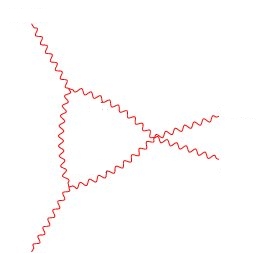}}&
\subfigure[fish topology]{\label{fig:edge-c}\includegraphics[scale=0.45]{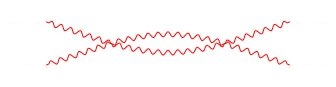}}
\end{tabular}
\caption[$\gamma\gamma\to\gamma\gamma$ process EW diagrams]
        {$\gamma\gamma\to\gamma\gamma$ process EW diagrams\label{EW.diagrams}}
\end{figure}

In the EW boson sector we have three types of diagrams to classify: box topologies, pinch topologies and fish topologies (shown in Figure~\ref{EW.diagrams}). There are three channels of each topologies (st-, su- and ut-channel as in QED) and we have $W^+$, $W^-$, $\phi^+$, $\phi^-$ and $X^+$, $X^-$ (bosons and ghosts) as internal particles in $R_\xi$-gauge theory.

As in fermionic part we can choose only positive or negative charge bosons and $X^+$, $X^-$ ghosts to appear as loop particles and multiply the result by factor 2 to dissmiss the double counting diagrams, which differ from another only by the orientation of the loop charge flow.

So, we have three structures (3 channels) in box type of diagrams, twelve structures (3 channels by 4 corresponding pinches) in pinch type and six structures (3 channels by 2 corresponding combinations of propagators --- direct and crossed) in fish type of diagrams --- each of ones is a sum of the appropriate sets of loop particles diagrams.

The full CA of given process for off-shell photons ($p_i\epsilon_i\neq 0$) with corresponding combinatorical factors can be written as sum of bosonic part minus fermionic and ghost part:
\bq
\begin{tabular}{m{14mm}m{16mm}m{18mm}m{16mm}m{18mm}m{16mm}m{18mm}m{2mm}}
${\cal A}_{\gamma\gamma\to\gamma\gamma}=$&
$+2\times\Big[\hspace{2mm}\sum\nolimits$&
\includegraphics[scale=0.2]{diagram_ew_box.jpg}&
$+\hspace{4mm}\frac{1}{2}\times\sum\nolimits$&
\includegraphics[scale=0.2]{diagram_ew_pinch.jpg}&
$+\hspace{4mm}\frac{1}{4}\times\sum\nolimits$&
\includegraphics[scale=0.2]{diagram_ew_fish.jpg}&
$\Big]$\\
&
$-2\times\Big[\hspace{2mm}\sum\nolimits$&
\includegraphics[scale=0.2]{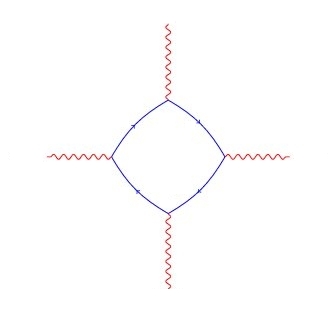}&
$\hspace{4mm}\Big]$&
$-2\times\Big[\hspace{2mm}\sum\nolimits$&
\includegraphics[scale=0.2]{diagram_qed.jpg}&
$\hspace{4mm}\Big]$&
\end{tabular}
\eq

In terms of Lorenz-structures we have:
\bq
{\cal A}_{\gamma\gamma\to\gamma\gamma}
  =\sum\limits_{i=1}^{43}\Big[{\cal F}_{i}^{\rm bosons}\left(s\,,t\,,u\right) + {\cal F}_{i}^{\rm fermions}\left(s\,,t\,,u\right)\Big] T_{i}^{\alpha\beta\mu\nu}.
\eq

The ${\cal F}_{i}$ are normilized by corresponding factors for fermion and boson parts:
\bqa
C^{\rm fermions} &=& 8 \alpha^2 Q_f^4 N_c\,,\nll
C^{\rm bosons} &=& 12 \alpha^2 \,,
\label{Cfactors}
\eqa
where $\alpha$ is the fine structure constant, $Q_f$ is the fraction of charge of loop fermion in units of electron charge $e$, $N_c$ is the number of colours for given fermion, $T_{i}^{\alpha\beta\mu\nu}$ are tensors defined with an aid of auxiliary strings $\tau_{j}$ presented in the Appendix section~\ref{sbasis}. The off-shell process contains 43 basis elements, but for the on-shell real photons we need only first 10 structures.

Thus, one gets a minimal number of tensor structures of the CA. It can be written in an explicit form with an aid of scalar ${\cal F}_{i}$. All masses and other parameters dependences are included into these ${\cal F}_{i}$. It is important to emphasize that each basis elements $T^{\alpha\beta\mu\nu}_{i},\,i=1\div43$ of tensor structure with Lorenz indices is four times transversal with respect to each external photon:
\bqa
T^{\alpha\beta\mu\nu}_{i}p_{\alpha}=
T^{\alpha\beta\mu\nu}_{i}p_{\beta}=
T^{\alpha\beta\mu\nu}_{i}p_{\mu}=
T^{\alpha\beta\mu\nu}_{i}p_{\nu}=0\,.
\label{WI}
\eqa

${\cal F}_{i}$ are scalar coefficients in front of basis structures of the CA --- projections of CA to complete basis expressions $T_{i}^{\alpha\beta\mu\nu}$. They are presented as some combinations of scalar Passarino--Veltman functions $A_0$, $B_0$, $C_0$, $D_0$~\cite{Passarino:1979} and depended on invariants $s\,,t\,,u$ and also on fermions and bosons masses. They do not contain UV poles.

The number of terms in ${\cal F}_{i}$ equals to thousands in the case of non-zero mass of the loop particles, but this number reduces greatly for zero loop fermion mass. Full answer for ${\cal F}_{i}$ one can find in the computer system {\tt SANC}. The client part is available on servers~\cite{SANC:2006}.


\subsection{Helicity amplitudes \label{HA}}

In {\tt SANC} we use helicity amplitudes approach~\cite{Andonov:2006,VW:1996}.

In the expression for CA as one can see in subsection~\ref{CA} one has tensor structures and a set of scalar ${\cal F}_{i}$. To calculate an observable quantity, such as cross section, one needs to make amplitude square, calculate, in general, products of Dirac spinors and contract Lorenz indices with polarization vector. In the standard approach making amplitude square one gets squares for each diagram and their interferences. This leads to a huge number of terms.

In the helicity amplitudes approach we also derive tensor structure and ${\cal F}_{i}$. But the next step is a projection to helicity basis and as a result one gets a set of non-interfering amplitudes, since all of them are characterized by different set of helicity quantum numbers. In this way we can distinguish calculations of Dirac spinors, if they are needed, and contraction of Lorenz indices from calculations of ${\cal F}_{i}$. We can do this before making square of amplitude. So we get a profit on calculation time (less amount of terms due to zero interference) and also more clear step-by-step control.

For the process $\gamma\gamma\to\gamma\gamma$ one gets:
\bqa
{\cal A}_{\gamma\gamma\to\gamma\gamma}&=& \sum_{\rm{spins}}\Big[C^{\rm bosons}\times{\cal H}_{\rm{spins}}^{\rm bosons} + C^{\rm fermions}\times{\cal H}_{\rm{spins}}^{\rm fermions}\Big]\,,\nll
\left|{\cal A}_{\gamma\gamma\to\gamma\gamma}\right|^2&=& \sum_{\rm{spins}}\Big[C^2_{\rm bosons}|{\cal H}_{\rm{spins}}^{\rm bosons}|^2 + C^2_{\rm fermions}|{\cal H}_{\rm{spins}}^{\rm fermions}|^2\nll
&&+ C^{\rm bosons}C^{\rm fermions}\Big({\cal H}_{\rm{spins}}^{\rm *bosons}\times{\cal H}_{\rm{spins}}^{\rm fermions}+{\cal H}_{\rm{spins}}^{\rm bosons}\times{\cal H}_{\rm{spins}}^{\rm *fermions}\Big)\Big]\,.
\eqa

Note, the total number of HAs for this process is equal to 16. This corresponds to different combinations of external particles spin projections. For $\gamma\gamma\to\gamma\gamma$ processes there are 4 photons with two polarizations $'+'$ and $'-'$, so the total number is $2\cdot2\cdot2\cdot2=16$. Helicity amplitudes are scalar expressions with factors equal to $C^{\rm fermions}$ for fermions and $C^{\rm bosons}$ for bosons~(\ref{Cfactors}).

For bosons part we have:
\bqa
{\cal H}^{\rm bosons}_{++++}={\cal H}^{\rm bosons}_{----}&=& -1+\frac{u-t}{s}\Big[B_{0}(u;M_W,M_W)-B_{0}(t;M_W,M_W)\Big]\nll
&&+\Big(\frac{4 M_W^2}{s}+2(\frac{tu}{s^2}-\frac{4}{3})\Big)\Big(u C_0(0,0,u;M_W,M_W,M_W) + t C_0(0,0,t;M_W,M_W,M_W)\Big)\nll
&&-\Big(2 M_W^2 s (\frac{M_W^2}{s} - \frac{4}{3}) + \frac{2s^2}{3}\Big)\Big( D_0(0,0,0,0,s,t;M_W,M_W,M_W,M_W)\nll
&&+D_0(0,0,0,0,s,u;M_W,M_W,M_W,M_W)+D_0(0,0,0,0,u,t;M_W,M_W,M_W,M_W)\Big)\nll
&&- t u \Big( \frac{4 M_W^2}{s}+\frac{t u}{s^2}-\frac{4}{3}\Big) D_0(0,0,0,0,u,t;M_W,M_W,M_W,M_W)\,, \nll
{\cal H}^{\rm bosons}_{+++-}={\cal H}^{\rm bosons}_{++-+}&=& {\cal H}^{\rm bosons}_{+-++}={\cal H}^{\rm bosons}_{-+++} ={\cal H}^{\rm bosons}_{---+}={\cal H}^{\rm bosons}_{--+-}= \nll
{\cal H}^{\rm bosons}_{-+--}={\cal H}^{\rm bosons}_{+---}&=& 1-M_W^2\Big(s^2+t^2+u^2\Big)\Big(\frac{1}{ut} C_0(0,0,s;M_W,M_W,M_W)\nll
&&+\frac{1}{su}C_0(0,0,t;M_W,M_W,M_W)+\frac{1}{st}C_0(0,0,u;M_W,M_W,M_W)\Big)-M_W^2\Big(\nll
&&(2M_W^2+\frac{st}{u})D_0(0,0,0,0,s,t;M_W,M_W,M_W,M_W)\nll
&&+(2M_W^2+\frac{su}{t})D_0(0,0,0,0,s,u;M_W,M_W,M_W,M_W)\nll
&&+(2M_W^2+\frac{ut}{s})D_0(0,0,0,0,u,t;M_W,M_W,M_W,M_W)\Big)\,,\nll 
{\cal H}^{\rm bosons}_{+--+} = {\cal H}^{\rm bosons}_{-++-} &=& -1+\frac{s-t}{u}\Big[B_{0}(s;M_W,M_W)-B_{0}(t;M_W,M_W)\Big]\nll
&&+\Big(\frac{4 M_W^2}{u}+2(\frac{ts}{u^2}-\frac{4}{3})\Big)\Big(s C_0(0,0,s;M_W,M_W,M_W) + t C_0(0,0,t;M_W,M_W,M_W)\Big)\nll
&&-\Big(2 M_W^2 u (\frac{M_W^2}{u} - \frac{4}{3}) + \frac{2u^2}{3}\Big)\Big( D_0(0,0,0,0,u,t;M_W,M_W,M_W,M_W)\nll
&&+D_0(0,0,0,0,s,u;M_W,M_W,M_W,M_W)+D_0(0,0,0,0,s,t;M_W,M_W,M_W,M_W)\Big)\nll
&&- t s \Big( \frac{4 M_W^2}{u}+\frac{t s}{u^2}-\frac{4}{3}\Big) D_0(0,0,0,0,s,t;M_W,M_W,M_W,M_W)\,, \nll
{\cal H}^{\rm bosons}_{+-+-} = {\cal H}^{\rm bosons}_{-+-+} &=& -1+\frac{u-s}{t}\Big[B_{0}(u;M_W,M_W)-B_{0}(s;M_W,M_W)\Big]\nll
&&+\Big(\frac{4 M_W^2}{t}+2(\frac{su}{t^2}-\frac{4}{3})\Big)\Big(u C_0(0,0,u;M_W,M_W,M_W) + s C_0(0,0,s;M_W,M_W,M_W)\Big)\nll
&&-\Big(2 M_W^2 t (\frac{M_W^2}{t} - \frac{4}{3}) + \frac{2t^2}{3}\Big)\Big( D_0(0,0,0,0,s,t;M_W,M_W,M_W,M_W)\nll
&&+D_0(0,0,0,0,u,t;M_W,M_W,M_W,M_W)+D_0(0,0,0,0,s,u;M_W,M_W,M_W,M_W)\Big)\nll
&&- s u \Big( \frac{4 M_W^2}{t}+\frac{s u}{t^2}-\frac{4}{3}\Big) D_0(0,0,0,0,s,u;M_W,M_W,M_W,M_W)\,, \nll
{\cal H}^{\rm bosons}_{++--}={\cal H}^{\rm bosons}_{--++}&=& 1-2M_W^4\Big( D_0(0,0,0,0,u,t;M_W,M_W,M_W,M_W)+\nll
&&D_0(0,0,0,0,s,u;M_W,M_W,M_W,M_W)+D_0(0,0,0,0,s,t;M_W,M_W,M_W,M_W)\Big).
\eqa

The analytical answers for scalar Passarino--Veltman functions $A_0$, $B_0$, $C_0$, $D_0$~\cite{Passarino:1979} are presented in corresponding Appendix sections.

For fermions part we have similar to bosons result with corresponding loop particles masses, but with opposite sign and there is a difference in helicities conserving part of amplitude ($++++$ and $+-+-$ permutations), the non-conserving helicities ($++--$) are the same:
\bqa
{\cal H}^{\rm fermions}_{++++}={\cal H}^{\rm fermions}_{----}&=& 1-\frac{u-t}{s}\Big[B_{0}(u;m_f,m_f)-B_{0}(t;m_f,m_f)\Big]\nll
&&-\Big(\frac{4 m_f^2}{s}+2(\frac{tu}{s^2}-\frac{1}{2})\Big)\Big(u C_0(0,0,u;m_f,m_f,m_f) + t C_0(0,0,t;m_f,m_f,m_f)\Big)\nll
&&+2 m_f^2 s \Big(\frac{m_f^2}{s} - \frac{1}{2}\Big) \Big( D_0(0,0,0,0,s,t;m_f,m_f,m_f,m_f)\nll
&&+D_0(0,0,0,0,s,u;m_f,m_f,m_f,m_f)+D_0(0,0,0,0,u,t;m_f,m_f,m_f,m_f)\Big)\nll
&&+ t u \Big( \frac{4 m_f^2}{s}+\frac{t u}{s^2}-\frac{1}{2}\Big) D_0(0,0,0,0,u,t;m_f,m_f,m_f,m_f)\,, \nll
{\cal H}^{\rm fermions}_{+++-}={\cal H}^{\rm fermions}_{++-+}&=& {\cal H}^{\rm fermions}_{+-++}={\cal H}^{\rm fermions}_{-+++} ={\cal H}^{\rm fermions}_{---+}={\cal H}^{\rm fermions}_{--+-}= \nll
{\cal H}^{\rm fermions}_{-+--}={\cal H}^{\rm fermions}_{+---}&=& -1+m_f^2\Big(s^2+t^2+u^2\Big)\Big(\frac{1}{ut} C_0(0,0,s;m_f,m_f,m_f)\nll
&&+\frac{1}{su}C_0(0,0,t;m_f,m_f,m_f)+\frac{1}{st}C_0(0,0,u;m_f,m_f,m_f)\Big)+m_f^2\Big(\nll
&&(2m_f^2+\frac{st}{u})D_0(0,0,0,0,s,t;m_f,m_f,m_f,m_f)\nll
&&+(2m_f^2+\frac{su}{t})D_0(0,0,0,0,s,u;m_f,m_f,m_f,m_f)\nll
&&+(2m_f^2+\frac{ut}{s})D_0(0,0,0,0,u,t;m_f,m_f,m_f,m_f)\Big)\,,\nll 
{\cal H}^{\rm fermions}_{+--+} = {\cal H}^{\rm fermions}_{-++-} &=& 1-\frac{s-t}{u}\Big[B_{0}(s;m_f,m_f)-B_{0}(t;m_f,m_f)\Big]\nll
&&-\Big(\frac{4 m_f^2}{u}+2(\frac{ts}{u^2}-\frac{1}{2})\Big)\Big(s C_0(0,0,s;m_f,m_f,m_f) + t C_0(0,0,t;m_f,m_f,m_f)\Big)\nll
&&+2 m_f^2 u \Big(\frac{m_f^2}{u} - \frac{1}{2}\Big)\Big( D_0(0,0,0,0,u,t;m_f,m_f,m_f,m_f)\nll
&&+D_0(0,0,0,0,s,u;m_f,m_f,m_f,m_f)+D_0(0,0,0,0,s,t;m_f,m_f,m_f,m_f)\Big)\nll
&&+ t s \Big( \frac{4 m_f^2}{u}+\frac{t s}{u^2}-\frac{1}{2}\Big) D_0(0,0,0,0,s,t;m_f,m_f,m_f,m_f)\,, \nll
{\cal H}^{\rm fermions}_{+-+-} = {\cal H}^{\rm fermions}_{-+-+} &=& 1-\frac{u-s}{t}\Big[B_{0}(u;m_f,m_f)-B_{0}(s;m_f,m_f)\Big]\nll
&&-\Big(\frac{4 m_f^2}{t}+2(\frac{su}{t^2}-\frac{1}{2})\Big)\Big(u C_0(0,0,u;m_f,m_f,m_f) + s C_0(0,0,s;m_f,m_f,m_f)\Big)\nll
&&+2 m_f^2 t \Big(\frac{m_f^2}{t} - \frac{1}{2}\Big)\Big( D_0(0,0,0,0,s,t;m_f,m_f,m_f,m_f)\nll
&&+D_0(0,0,0,0,u,t;m_f,m_f,m_f,m_f)+D_0(0,0,0,0,s,u;m_f,m_f,m_f,m_f)\Big)\nll
&&+ s u \Big( \frac{4 m_f^2}{t}+\frac{s u}{t^2}-\frac{1}{2}\Big) D_0(0,0,0,0,s,u;m_f,m_f,m_f,m_f)\,, \nll
{\cal H}^{\rm fermions}_{++--}={\cal H}^{\rm fermions}_{--++}&=& -1+2m_f^4\Big( D_0(0,0,0,0,u,t;m_f,m_f,m_f,m_f)+\nll
&&D_0(0,0,0,0,s,u;m_f,m_f,m_f,m_f)+D_0(0,0,0,0,s,t;m_f,m_f,m_f,m_f)\Big).
\eqa

In massive case we observe five independent HAs, while in the case of zero loop fermion mass ($m_f=0$) one gets only four independent HAs which are very compact:
\bqa
&&\lk\lk\lk
{\ds {\cal H}^{\rm fermions}_{++++} = {\cal H}^{\rm fermions}_{----}=
-1+\left(\frac{t-u}{s}\right)\left(l_u-l_t\right)
-\left(\frac{1}{2}-\frac{ut}{s^2}\right) \left( \left(l_u - l_t\right)^2 + \pi^2 \right),}\nll
&&\lk\lk\lk
{\ds {\cal H}^{\rm fermions}_{+--+} = {\cal H}^{\rm fermions}_{-++-} = -1-i\pi\bigg(\frac{t-s}{u}\bigg)\,
-\Bigg[\left(1+i\pi\right)\bigg(\frac{t-s}{u}\bigg)\,+2i\pi\bigg(\frac{t}{u}\bigg)^2\Bigg] l_t\,
-\bigg(\frac{1}{2}-\frac{st}{u^2}\bigg) l^2_t \,,}\nll
&&\lk\lk\lk
{\ds{\cal H}^{\rm fermions}_{+-+-} = {\cal H}^{\rm fermions}_{-+-+} =
-1-i\pi\bigg(\frac{u-s}{t}\bigg)
-\Bigg[\left(1+i\pi\right)\bigg(\frac{u-s}{t}\bigg)+2i\pi\bigg(\frac{u}{t}\bigg)^2\Bigg] l_u
-\bigg(\frac{1}{2}-\frac{su}{t^2}\bigg) l^2_u \,.}
\eqa
where
\bqa
l_t=\ln\left(-\frac{t}{s}\right),\qquad l_u=\ln\left(-\frac{u}{s}\right).
\eqa

All the others fermionic HAs are equal to $+1$.

There are relations among helicity HAs due to C,P,T-invariance. Moreover, there is another relation due to crossing symmetry:
\bqa
{\cal H}_{+--+}\left(s,t,u\right)={\cal H}_{+-+-}\left(s,u,t\right)\,,
\eqa
but this fact does not mean reducing the number of independent HAs.


\section{Process $\gamma\gamma\to\gamma\gamma$ in {\tt SANC}\label{tree}}
\subsection{{\tt SANC} process tree}

In this section we briefly describe analytic modules relevant for $\gamma\gamma\to\gamma\gamma$.

For boxes the {\tt SANC} idea of precomputation becomes vitally important~\cite{Andonov:2006}. Calculation of some boxes for some particular processes takes so much time that an external user should refrain from repeating precomputation in the {\tt SANC} system. Furthermore, the richness of boxes requires a classification. Depending on the type of external lines ($f$ for fermions and $b$ for bosons), we distinguish three large classes of boxes: $ffff$, $ffbb$ and $bbbb$.

\begin{figure}[htp]
\begin{tabular}{m{8cm}m{8cm}}
\subfigure[QED tree]{\label{fig:edge-a}\includegraphics[scale=0.85]{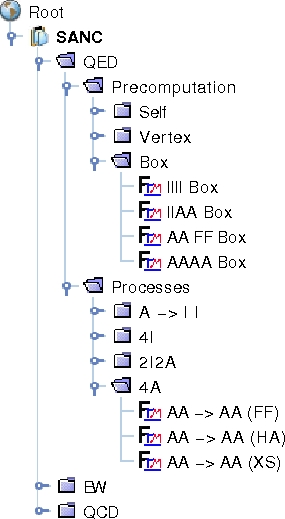}}&
\subfigure[EW tree]{\label{fig:edge-b}\includegraphics[scale=0.58]{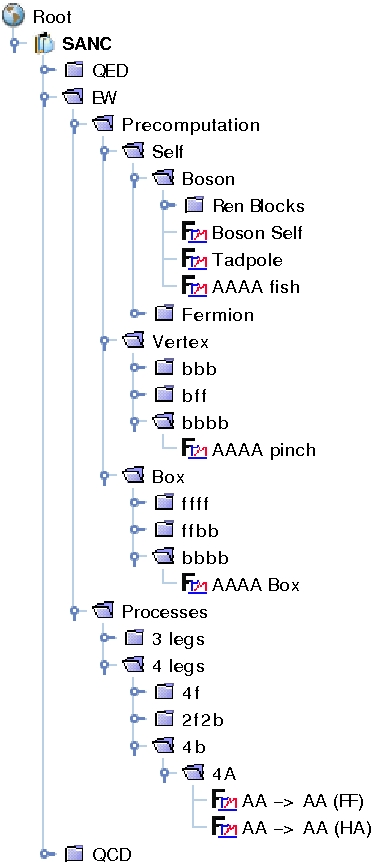}}
\end{tabular}
\caption[$\gamma\gamma\to\gamma\gamma$ in the {\tt SANC} process tree]
        {$\gamma\gamma\to\gamma\gamma$ in the {\tt SANC} process tree\label{tree.SANC}}
\end{figure}

The sum of contributions of fermionic loop boxes form a gauge-invariant and UV-finite subset, which is a consequence of Ward Identity \eqn{WI}. It is true for sum of bosonic contribution too. So we can distinguish QED and EW part of the process in the analitycal calculations.

The precomputation file {\tt AAAA Box} (see the {\tt SANC} process tree in Figure \ref{tree.SANC} from {\tt SANC} client \cite{SANC:2006}) contains the sequence of procedures for calculation of the covariant amplitude. At this step we suppose, that all momenta are incoming (denoted by $p$'s) and photons are not on-mass-shell. Therefore, these results can be used for other processes which need these parts as building blocks.

When we implement the process $\gamma\gamma\to\gamma\gamma$ (see {\tt 4A} QED or EW Processes branch), we use this building block several times by replacing incoming momenta $p$'s by corresponding kinematical momenta with the right signs, and calculate ${\cal F}_{i}$ by the module {\tt AA->AA (FF)}, then helicity amplitudes by the module {\tt AA->AA (HA)} and finally --- the analytic expression for differential and total process cross section (XS) by the module {\tt AA->AA (XS)} for the QED process. For the EW part we use numerical evaluation to get XS.
 
\subsection{Stand-alone SANC module\label{module}}
Now let us introduce the {\tt SANC} modules packages concept.

In {\tt SANC} system one has an opportunity of exporting the analytical results for numerical evolution~\cite{Andonov:2006}. Moreover, there are tools for checking the implementation of these Standard {\tt SANC FORTRAN} modules ({\tt SSFM})~\cite{Modules:2009} --- the integrator of the process, based on the Vegas algorithm~\cite{Vegas}.

For light-by-light scattering process the {\tt SSFM} are included in {\rm sanc\_4A\_v1.00} package, which is available for download from~\cite{SANC:2006}. The numerical results are cross-section distributions from next section. Here we present the technical description of this package --- some main flags and the options.

\begin{small}
\begin{figure}[htp]
\begin{center}
\includegraphics[scale=0.45]{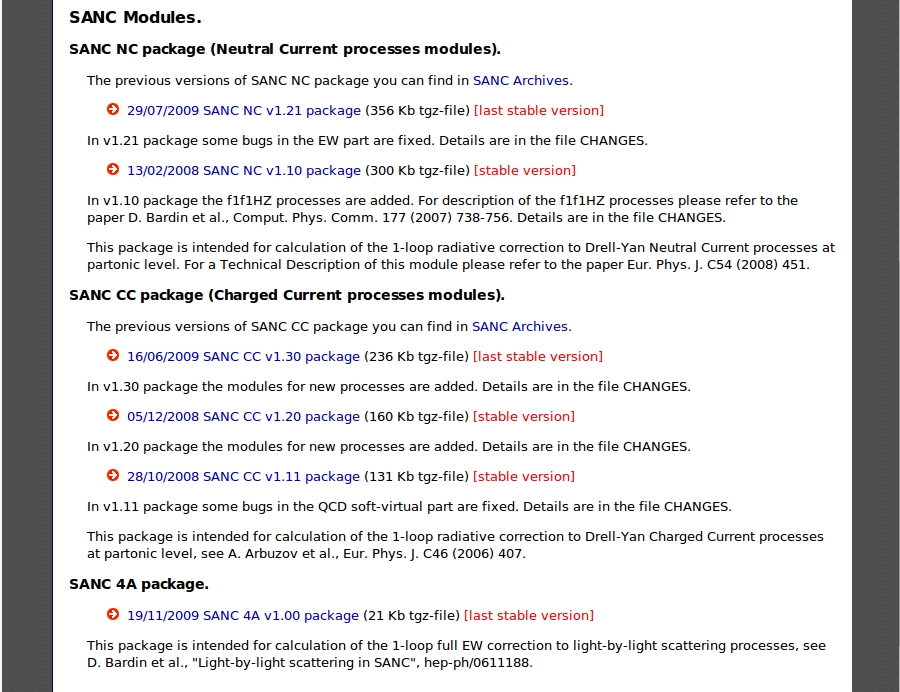}
\caption[{\tt SANC} modules download web site~\cite{SANC:2006}]
        {{\tt SANC} modules download web site~\cite{SANC:2006}}
\end{center}
\end{figure}
\end{small}

The package files:
\begin{itemize}
\item {\rm bbbb\_main\_11\_11.F} --- the main file;
\item {\rm bbbb\_ha\_11\_11.F} --- the HA from {\tt SANC} system;
\item {\rm *\_r16.f} --- the library of special functions and algorithms;
\item {\rm *\_input.h} --- the set of various setups of input parameters;
\item {\rm README, INSTALL} and other instructions files.
\end{itemize}

In {\rm README} and {\rm INSTALL} files one can find instructions how to use the package. The main options one can change in {\rm bbbb\_main\_11\_11.F} --- the main file.

There are some main flags and the options with next variants.

\begin{itemize}
\item {\bf iha(I)} --- choice of helicities sum in cross-section:\\
{\bf I=0}  total helicities sum\\
{\bf I=1}  $++$ helicities sum\\
{\bf I=2}  $+-$ helicities sum
\item {\bf iqed(I)} --- choice of calculations for QED corrections:\\
{\bf I=0}  without QED corrections\\
{\bf I=1}  with QED corrections
\item {\bf iew(I)} --- choice of calculations for EW corrections:\\
{\bf I=0}  without EW corrections\\
{\bf I=1}  with EW corrections
\item {\bf gfscheme(I)} --- choice of the EW scheme:\\
{\bf I=0}  $\alpha_0$ calculation scheme\\
{\bf I=1}  $G_F$ scheme\\
{\bf I=2}  ${G'}_F$ scheme, a test option, when $\alpha_0$ is replaced by $\alpha_{G_F}=\sqrt{2}G_F M^2_W \left(1-M^2_W / M^2_Z \right) / \pi$ 
\item {\bf isetup(I)} --- choice of the setup:\\
{\bf I=0}  Standard SANC input\\
{\bf I=1}  Les Houches Workshop (2005)\\
{\bf I=2}  Tevatron-for-LHC Workshop (2006)\\
{\bf I=3}  Custom setup~(\ref{setup})
\item {\bf start(I)} --- choice of start point for 4 order logarithmic scale of $\sqrt{s}$:\\
{\bf I=1q-5}  logarithmic scale $\sqrt{s}$ from $0.1~{\rm MeV}$ to $1~{\rm GeV}$\\
{\bf I=1q-1}  logarithmic scale $\sqrt{s}$ from $1~{\rm GeV}$ to $10^4~{\rm GeV}$   
\end{itemize} 

To get full EW answer with interfarence one should set {\bf iqed=1} and {\bf iew=1}. 

\subsection{Results and comparison\label{results}}

To test our analytical results we calculate final answer for the total cross section for QED part in the massless limit after substitutions of helicity amplitudes and angular integration:
\bq
\sigma^{\rm QED}_{\gamma\gamma\to\gamma\gamma} = \frac{e^8}{2\pi\omega^2} \left(\frac{108}{5} 
    + \frac{13}{2}\pi^2 - 8\pi^2\zeta(3) + \frac{148}{225}\pi^4 - 24\zeta(5) \right).
\eq

This result was compared with \cite{Ahiezer:1981} and the complete agreement was found. Also the limit of helicity amplitudes QED was compared separately with~\cite{Bern:2001} and again full agreement was observed. The massive expression of HA for QED and EW parts were compared with \cite{Jikia:1997} and \cite{Bohm:1994sf}.

\begin{small}
\begin{figure}[htp]
\begin{center}
\includegraphics[scale=1.0]{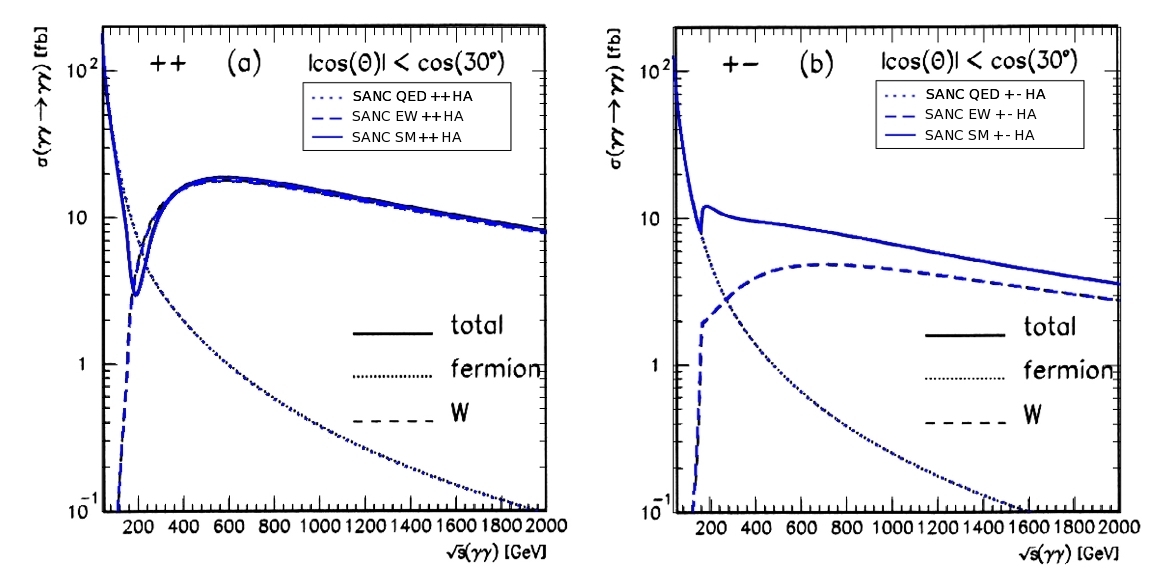}
\caption[$\gamma\gamma\to\gamma\gamma$ {\tt SANC} and \cite{Jikia:1997} cross-section ($++$ and $+-$) comparison]
        {$\gamma\gamma\to\gamma\gamma$ {\tt SANC} and \cite{Jikia:1997} cross-section ($++$ and $+-$) comparison\label{GeV.comparison}}
\end{center}
\end{figure}
\end{small}

\begin{small}
\begin{figure}[htp]
\begin{center}
\includegraphics[scale=0.59]{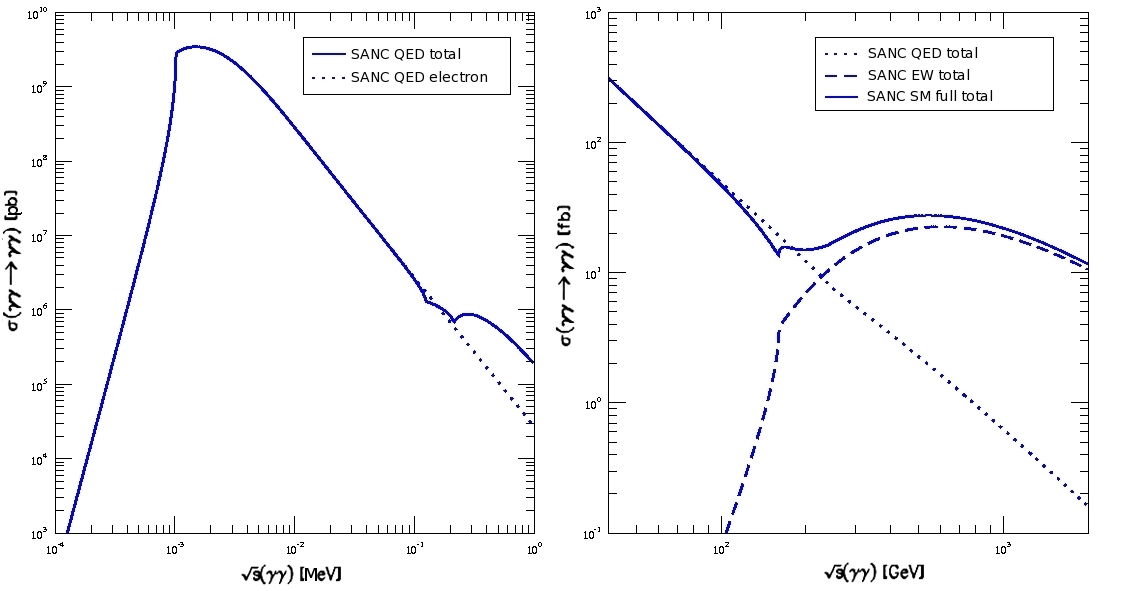}
\caption[$\gamma\gamma\to\gamma\gamma$ {\tt SANC} total cross-section]
        {$\gamma\gamma\to\gamma\gamma$ {\tt SANC} total cross-section\label{MeV.comparison}}
\end{center}
\end{figure}
\end{small}

The numerical evaluation of analytical expressions for QED, weak and full EW parts within {\tt SSFM} sanc\_4A\_v1.00 package for cross section were comparied with \cite{Jikia:1997} (see Figure~\ref{GeV.comparison}). The setup is: $\cos{\theta} < \cos{\pi/6}$ --- cuts; $M_W=80.22~{\rm GeV}$; $m_t=120~{\rm GeV}$; $\alpha=1/137.036$; other fermions are massless.

Also according to~\cite{Bohm:1994sf} the sum of helicities cross-sections were calculated for MeV and GeV regions (see Figure~\ref{MeV.comparison}) with custom setup:
\bqa
&\bullet& \alpha = 1/128;\nll
&\bullet& \cos{\theta} < \cos{\pi/6};\nll
&\bullet& M_W=80.22~{\rm GeV;}\nll
&\bullet& m_e = 0.51099892~{\rm MeV,}~~~m_\mu=0.105658369~{\rm GeV,}~~~m_\tau=1.77699~{\rm GeV;}\nll
&\bullet& m_u= 0.062~{\rm GeV,}~~~m_c= 1.50~{\rm GeV,}~~~m_t= 120.0~{\rm GeV;}\nll
&\bullet& m_d= 0.083~{\rm GeV,}~~~m_s= 0.215~{\rm GeV,}~~~m_b= 4.7~{\rm GeV.}
\label{setup}
\eqa

\section{Conclusions\label{conclusions}}
Let us summarise the results.

In this paper the helicities amplitudes expressions for light-by-light scattering in general (massive) and in limiting (massless) cases were listed for QED (fermionic) and weak (bosonic) parts.

We shortly described precomputation strategy of the {\tt SANC} system \cite{Andonov:2006} and the place of this process on the {\tt SANC} process tree.

The implementation of analitycal results and the {\tt SSFM} concept were described.

Additionally, we calculated and gave the answers for particular cases of $D_0$, $C_0$ and $B_0$ Passarino--Veltman (PV) functions for massive and massless loop particles, strings and basis for covariant amplitude.

The {\tt SSFM} sanc\_4A\_v1.00 package was builded and tested. Its numerical results were comparied with those existing in the literature. The package is available for download at web pages \cite{SANC:2006}. 

\section*{Acknowledgments \label{acknowledgements}}
\addcontentsline{toc}{section}{Acknowledgments}

\vspace*{5mm}
\noindent
The authors are grateful to S.~Bondarenko, V.~Kolesnikov and A.~Sapronov for useful discussions of numerical calculations and to A.~Arbuzov for providing us with useful references.
\clearpage

\def\thesection{\Roman{section}}
\setcounter{section}{0}
\section*{Appendix \label{appendix}}
\addcontentsline{toc}{section}{Appendix}
To obtain the cross section of the $\gamma\gamma\to\gamma\gamma$ process in an analytic form we have to compute in an explicit form the master integrals, $B_0$, $C_0$, $D_0$ --- scalar PV functions \cite{Passarino:1979}, \cite{Bardin:1999}  --- for a particular set of parameters. In $D_0$ and $C_0$ functions one can see collinear divergences, but the differential cross section is free of mass singularities which completely cancel in the sum of all terms. The $A_0$ and $B_0$ functions contain UV divergences, which cancel in the sum of box contributions. In the process of computation we face also a problem of the ``angular edge'' divergences, but they are not physical and cancel completely.

\section{$B_0$ function}
The $B_0$ function for $\gamma\gamma\to\gamma\gamma$ process reads:

\bqa
B_0\left(Q^2;M,M\right)&=&\frac{1}{\bar\epsilon}+2-\ln\left(\frac{M^2}{\mu^2}\right)
              -\beta\ln\left(\frac{\beta+1}{\beta-1}\right)\,,
\eqa
where $$\beta^2=1-\frac{4\widetilde{M}^2}{Q^2}\,,$$ and

\bqa
\widetilde{M}^2=M^2-{i}\epsilon\,.
\eqa

In zero limit of fermion mass: $M\to0$, one gets:
\bqa
B_0\left(s;M,M\right)&=&\frac{1}{\bar\epsilon}+2-\left[\ln\left(\frac{s}{\mu^2}\right)-{i}\pi\right],
\nll
B_0\left(t;M,M\right)&=&\frac{1}{\bar\epsilon}+2-\left[l_t+\ln\left(\frac{s}{\mu^2}\right)\right],
\nll
B_0\left(u;M,M\right)&=&\frac{1}{\bar\epsilon}+2-\left[l_u+\ln\left(\frac{s}{\mu^2}\right)\right].
\eqa

\section{$C_0$ function}
The $C_0$ function is:
\bqa
C_0\left(0,0,Q^2;M,M,M\right)=
\int\limits_{0}^{1}dx\int\limits_{0}^{x}dy\left(Q^2y-Q^2xy+\widetilde{M}^2\right)^{-1},
\eqa
\noindent

After calculations:
\bqa
C_0\left(0,0,Q^2;M,M,M\right)=-\frac{1}{Q^2}\left[\Litwo\left(\frac{1}{x_1}\right)
                                                 +\Litwo\left(\frac{1}{x_2}\right)\right],
\eqa
where
\bqa
x_{1,2}=\frac{1}{2}\left(1\pm\beta\right).
\eqa
\noindent

For $M\to0$:
\bqa
C_0\left(0,0,s;M,M,M\right)&=&-\frac{1}{2s}\left[\ln\left(\frac{M^2}{s}\right)+{i}\pi\right]^2,
\nll
C_0\left(0,0,t;M,M,M\right)&=&-\frac{1}{2t}\left[\ln\left(\frac{M^2}{s}\right)-l_t\right]^2,
\nll
C_0\left(0,0,u;M,M,M\right)&=&-\frac{1}{2u}\left[\ln\left(\frac{M^2}{s}\right)-l_u\right]^2.
\eqa

\section{$D_0$ function}
The $D_0$ function looks like:
\bqa
D_0\left(0,0,0,0,Q^2,P^2;M,M,M,M \right)=
\int\limits_{0}^{1}dx \int\limits_{0}^{x}dy 
       \int\limits_{0}^{y}dz \left(\left(Q^2+P^2\right)xz-P^2xy-Q^2yz+P^2y-P^2z+\widetilde{M}^2\right)^{-2}
\eqa

After lengthy calculations:
\bqa
&& D_0\left(0,0,0,0,Q^2,P^2;M,M,M,M\right) = -\frac{2}{stA_3}\Biggl\{\nll 
&& \Litwo\left(\frac{1+A_3}{A_1+A_3}\right)-\Litwo\left(\frac{A_{33}}{A_{13}}\right)
  +\Litwo\left(\frac{1+A_3}{A_2+A_3}\right)-\Litwo\left(\frac{A_{33}}{A_{23}}\right)\nll
&&-\Litwo\left(-\frac{A_{33}}{A_1+A_3}\right)-\Litwo\left(-\frac{A_{33}}{A_2+A_3}\right)
  -\Litwo\left(-\frac{A_{13}}{1+A_3}\right)-\Litwo\left(-\frac{A_{23}}{1+A_3}\right)\nll
&&-\frac{1}{2}\ln^2\left(\frac{A_{13}}{1+A_3}\right)-\frac{1}{2}\ln^2\left(\frac{A_{23}}{1+A_3}\right)-2\zeta(2)+2i\pi\ln\left(1+\frac{1}{A_3}\right)\theta(-Im(A_1+A_3))\Biggl\}\,,
\eqa
where
\bqa
 A_1=\sqrt{1-\frac{4\widetilde{M}^2}{Q^2}}\,,~~
 A_2=\sqrt{1-\frac{4\widetilde{M}^2}{P^2}}\,,~~
 A_3=\sqrt{1-\frac{4\widetilde{M}^2\left(Q^2+P^2\right)}{Q^2P^2}}\,,\nll
 A_{33}=\frac{4\widetilde{M}^2(Q^2+P^2)}{Q^2P^2(1+A_3)}\,,~~
 A_{23}=\frac{4\widetilde{M}^2}{Q^2(A_2+A_3)}\,,~~
 A_{13}=\frac{4\widetilde{M}^2}{P^2(A_1+A_3)}.
\eqa
$\theta\left(x\right)$ is a function of Heaviside.

For $M\to0$:
\bqa
\lk\lk
D_0\left(0,0,0,0,s,t;M,M,M,M\right) &\lk=\lk& \frac{2}{st}\left[ \ln^2\left(-\frac{M^2}{t}\right)
 + \ln\left(-\frac{M^2}{t}\right) l_t -\frac{\pi^2}{2}+i\pi \ln\left(-\frac{M^2}{t}\right) \right]\nll
\lk\lk
D_0\left(0,0,0,0,u,t;M,M,M,M\right) &\lk=\lk& \frac{2}{ut}\left[ \ln^2\left(\frac{M^2}{s}\right) 
 - \ln\left(\frac{M^2}{s}\right) \left(l_t+l_u\right) 
-\frac{\pi^2}{2}+l_t l_u \right],
\nll
\lk\lk
D_0\left(0,0,0,0,s,u;M,M,M,M\right) &\lk=\lk& \frac{2}{su}\left[ \ln^2\left(-\frac{M^2}{u}\right)
+ \ln\left(-\frac{M^2}{u}\right) l_u -\frac{\pi^2}{2}+i\pi \ln\left(-\frac{M^2}{u}\right) \right].
\eqa

\section{Strings and basis\label{sbasis}}
To obtain a compact form of structures of the amplitude we choose 14 auxiliary tensorial strings:
\bq
\begin{array}{llllllllll}
&\tau_{1}^{\alpha\beta} &\lk=\lk& p_{1\beta}p_{2\alpha}+\frac{1}{2}s\delta_{\alpha\beta}\,,
&\tau_{2}^{\mu\nu}      &\lk=\lk& p_{3\mu}p_{4\nu}+\frac{1}{2}s\delta_{\mu\nu}\,, 
&\tau_{3}^{\beta\nu}    &\lk=\lk& p_{2\nu}p_{3\beta}+\frac{1}{2}t\delta_{\beta\nu}\,,
\nll
&\tau_{4}^{\alpha\mu}   &\lk=\lk& p_{1\mu}p_{4\alpha}+\frac{1}{2}t\delta_{\alpha\mu}\,,
&\tau_{5}^{\alpha\nu}   &\lk=\lk& p_{1\nu}p_{3\alpha}+\frac{1}{2}u\delta_{\alpha\nu}\,,
&\tau_{6}^{\beta\mu}    &\lk=\lk& p_{4\beta}p_{2\mu}+\frac{1}{2}u\delta_{\beta\mu}\,,
\nll
&\tau_{7}^{\mu}         &\lk=\lk& p_{1\mu}-tu^{-1}p_{2\mu}\,,
&\tau_{8}^{\nu}         &\lk=\lk& p_{1\nu}-ut^{-1}p_{2\nu}\,,
&\tau_{9}^{\beta}       &\lk=\lk& p_{1\beta}-st^{-1}p_{3\beta}\,,
\nll
&\tau_{10}^{\alpha}     &\lk=\lk& p_{2\alpha}-su^{-1}p_{3\alpha}\,,
&\tau_{11}^{\mu}        &\lk=\lk& p_{4\mu}\,,~~\tau_{12}^{\nu}=p_{3\nu},
&\tau_{13}^{\beta}      &\lk=\lk& p_{2\beta}\,,~~\tau_{14}^{\alpha}=p_{1\alpha}.
\end{array}
\eq

The complete basis $T_{i}^{\alpha\beta\mu\nu}$ 
can be presented in a compact form with an aid of the auxiliary strings $\tau_{j}$:
\bq
\begin{array}{lllllllllllll}
&T^{\alpha\beta\mu\nu}_{1}  &\lk=\lk& \tau_{1}^{\alpha\beta}\tau_{2}^{\mu\nu},              
&T^{\alpha\beta\mu\nu}_{2}  &\lk=\lk& \tau_{3}^{\beta\nu}   \tau_{4}^{\alpha\mu},	      
&T^{\alpha\beta\mu\nu}_{3}  &\lk=\lk& \tau_{5}^{\alpha\nu}  \tau_{6}^{\beta\mu}, 	      
&T^{\alpha\beta\mu\nu}_{4}  &\lk=\lk& \tau_{1}^{\alpha\beta}\tau_{7}^{\mu}    \tau_{8}^{\nu},
\nll
&T^{\alpha\beta\mu\nu}_{5}  &\lk=\lk& \tau_{2}^{\mu\nu}     \tau_{9}^{\beta}  \tau_{10}^{\alpha}, 
&T^{\alpha\beta\mu\nu}_{6}  &\lk=\lk& \tau_{3}^{\beta\nu}   \tau_{7}^{\mu}    \tau_{10}^{\alpha},
&T^{\alpha\beta\mu\nu}_{7}  &\lk=\lk& \tau_{4}^{\alpha\mu}  \tau_{8}^{\nu}    \tau_{9}^{\beta},  
&T^{\alpha\beta\mu\nu}_{8}  &\lk=\lk& \tau_{5}^{\alpha\nu}  \tau_{7}^{\mu}    \tau_{9}^{\beta},  
\nll
&T^{\alpha\beta\mu\nu}_{9}  &\lk=\lk& \tau_{6}^{\beta\mu}   \tau_{8}^{\nu}    \tau_{10}^{\alpha},
&T^{\alpha\beta\mu\nu}_{10} &\lk=\lk& \tau_{7}^{\mu}        \tau_{8}^{\nu}    \tau_{9}^{\beta}     \tau_{10}^{\alpha},
&T^{\alpha\beta\mu\nu}_{11} &\lk=\lk& \tau_{1}^{\alpha\beta}\tau_{7}^{\mu}    \tau_{12}^{\nu},   
&T^{\alpha\beta\mu\nu}_{12} &\lk=\lk& \tau_{2}^{\mu\nu}     \tau_{9}^{\beta}  \tau_{14}^{\alpha},
\nll
&T^{\alpha\beta\mu\nu}_{13} &\lk=\lk& \tau_{3}^{\beta\nu}   \tau_{7}^{\mu}    \tau_{14}^{\alpha}, 
&T^{\alpha\beta\mu\nu}_{14} &\lk=\lk& \tau_{4}^{\alpha\mu}  \tau_{8}^{\nu}    \tau_{13}^{\beta},  
&T^{\alpha\beta\mu\nu}_{15} &\lk=\lk& \tau_{5}^{\alpha\nu}  \tau_{7}^{\mu}    \tau_{13}^{\beta},  
&T^{\alpha\beta\mu\nu}_{16} &\lk=\lk& \tau_{6}^{\beta\mu}   \tau_{8}^{\nu}    \tau_{14}^{\alpha}, 
\nll			       
&T^{\alpha\beta\mu\nu}_{17} &\lk=\lk& \tau_{1}^{\alpha\beta}\tau_{11}^{\mu}   \tau_{8}^{\nu},     
&T^{\alpha\beta\mu\nu}_{18} &\lk=\lk& \tau_{2}^{\mu\nu}     \tau_{13}^{\beta} \tau_{10}^{\alpha}, 
&T^{\alpha\beta\mu\nu}_{19} &\lk=\lk& \tau_{3}^{\beta\nu}   \tau_{11}^{\mu}   \tau_{10}^{\alpha}, 
&T^{\alpha\beta\mu\nu}_{20} &\lk=\lk& \tau_{4}^{\alpha\mu}  \tau_{12}^{\nu}   \tau_{9}^{\beta},   
\nll	
&T^{\alpha\beta\mu\nu}_{21} &\lk=\lk& \tau_{5}^{\alpha\nu}  \tau_{11}^{\mu}   \tau_{9}^{\beta},   
&T^{\alpha\beta\mu\nu}_{22} &\lk=\lk& \tau_{6}^{\beta\mu}   \tau_{12}^{\nu}   \tau_{10}^{\alpha}, 
&T^{\alpha\beta\mu\nu}_{23} &\lk=\lk& \tau_{1}^{\alpha\beta}\tau_{11}^{\mu}   \tau_{12}^{\nu},    
&T^{\alpha\beta\mu\nu}_{24} &\lk=\lk& \tau_{2}^{\mu\nu}     \tau_{13}^{\beta} \tau_{14}^{\alpha}, 
\nll	
&T^{\alpha\beta\mu\nu}_{25} &\lk=\lk& \tau_{3}^{\beta\nu}   \tau_{11}^{\mu}   \tau_{14}^{\alpha},                      
&T^{\alpha\beta\mu\nu}_{26} &\lk=\lk& \tau_{4}^{\alpha\mu}  \tau_{12}^{\nu}   \tau_{13}^{\beta},	              
&T^{\alpha\beta\mu\nu}_{27} &\lk=\lk& \tau_{5}^{\alpha\nu}  \tau_{11}^{\mu}   \tau_{13}^{\beta}, 	              
&T^{\alpha\beta\mu\nu}_{28} &\lk=\lk& \tau_{6}^{\beta\mu}   \tau_{12}^{\nu}   \tau_{14}^{\alpha},
\nll			       
&T^{\alpha\beta\mu\nu}_{29} &\lk=\lk& \tau_{7}^{\mu}        \tau_{8}^{\nu}    \tau_{13}^{\beta}    \tau_{14}^{\alpha}, 
&T^{\alpha\beta\mu\nu}_{30} &\lk=\lk& \tau_{7}^{\mu}        \tau_{9}^{\beta}  \tau_{12}^{\nu}      \tau_{14}^{\alpha}, 
&T^{\alpha\beta\mu\nu}_{31} &\lk=\lk& \tau_{7}^{\mu}        \tau_{10}^{\alpha}\tau_{12}^{\nu}      \tau_{13}^{\beta},  
&T^{\alpha\beta\mu\nu}_{32} &\lk=\lk& \tau_{8}^{\nu}        \tau_{9}^{\beta}  \tau_{11}^{\mu}      \tau_{14}^{\alpha}, 
\nll	
&T^{\alpha\beta\mu\nu}_{33} &\lk=\lk& \tau_{8}^{\nu}        \tau_{10}^{\alpha}\tau_{13}^{\beta}  \tau_{11}^{\mu},    
&T^{\alpha\beta\mu\nu}_{34} &\lk=\lk& \tau_{9}^{\beta}      \tau_{10}^{\alpha}\tau_{11}^{\mu}      \tau_{12}^{\nu},    
&T^{\alpha\beta\mu\nu}_{35} &\lk=\lk& \tau_{7}^{\mu}        \tau_{8}^{\nu}    \tau_{9}^{\beta}     \tau_{14}^{\alpha}, 
&T^{\alpha\beta\mu\nu}_{36} &\lk=\lk& \tau_{7}^{\mu}        \tau_{8}^{\nu}    \tau_{10}^{\alpha}   \tau_{13}^{\beta},  
\nll
&T^{\alpha\beta\mu\nu}_{37} &\lk=\lk& \tau_{7}^{\mu}        \tau_{9}^{\beta}  \tau_{10}^{\alpha}   \tau_{12}^{\nu},     
&T^{\alpha\beta\mu\nu}_{38} &\lk=\lk& \tau_{8}^{\nu}        \tau_{9}^{\beta}  \tau_{10}^{\alpha}   \tau_{11}^{\mu},     
&T^{\alpha\beta\mu\nu}_{39} &\lk=\lk& \tau_{11}^{\mu}       \tau_{12}^{\nu}   \tau_{13}^{\beta}    \tau_{10}^{\alpha},  
&T^{\alpha\beta\mu\nu}_{40} &\lk=\lk& \tau_{11}^{\mu}       \tau_{12}^{\nu}   \tau_{14}^{\alpha}   \tau_{9}^{\beta},    
\nll			       
&T^{\alpha\beta\mu\nu}_{41} &\lk=\lk& \tau_{11}^{\mu}       \tau_{13}^{\beta} \tau_{14}^{\alpha}   \tau_{8}^{\nu},
&T^{\alpha\beta\mu\nu}_{42} &\lk=\lk& \tau_{12}^{\nu}       \tau_{13}^{\beta} \tau_{14}^{\alpha}   \tau_{7}^{\mu},
&T^{\alpha\beta\mu\nu}_{43} &\lk=\lk& \tau_{11}^{\mu}       \tau_{12}^{\nu}   \tau_{13}^{\beta}    \tau_{14}^{\alpha} .
&                           &\lk \lk&
\end{array}
\label{basis}
\eq

\def\href#1#2{#2}
\addcontentsline{toc}{section}{References}

\begingroup\begin{thebibliography}{10}

\bibitem{Andonov:2006}
A.~Andonov {\it et al.}, 
CPC 174 (2006), p.481-517.

\bibitem{Modules:2009}
A.~Andonov {\it et al.},
CPC 181 (2009), p.305-312.

\bibitem{SANC:2006}
WWW: http://sanc.jinr.ru, http://pcphsanc.cern.ch

\bibitem{Euler:1936}
H.~Euler, W.~Heisenberg, "Quantum Physics" (1936), p.98.

\bibitem{Karplus:1951}
R.~Karplus, M.~Neuman,
Phys. Rev., v.83 (1951) n.4, p.776.

\bibitem{Ahiezer:1981}
A.~Ahiezer, V.~Berestecky, "Quantum Electrodynamic", p.375, 4 edition, 1981.

\bibitem{Jikia:1993}
G.~Jikia and A.~Tkabladze,
"Photon-Photon Scattering at the Photon Linear Collider",
hep-ph/9312228

\bibitem{Jikia:1997}
G.~Jikia,
"Electroweak gauge boson production at $\gamma\gamma$ collider",
hep-ph/9710459

\bibitem{Bohm:1994sf}
M.~Bohm, R.~Schuster,
Z. Phys. C 63 (1994), p.219.

\bibitem{Diakonidis:2006}
Th.~Diakonidis {\it et al.},
"A FORTRAN code for $\gamma\gamma\to ZZ$ in SM and MSSM",
hep-ph/0610085.

\bibitem{Bern:2001}
Z.~Bern {\it et al.}, 
"QCD and QED Corrections to Light-by-Light Scattering",
hep-ph/0109079.

\bibitem{VW:1996}
R.~Vega, J.~Wudka,
Phys. Rev. D 53 (1996), p.5286-5292.

\bibitem{Passarino:1979}
G.~Passarino, M.~Veltman, Nuclear Physics, B 166 (1979), p.151.

\bibitem{Bardin:1999}
D.~Bardin, G.~Passarino, "The Standard Model in the making", Oxford, 1999

\bibitem{Bardin:2006}
D.~Bardin {\it et al.},
"Light-by-light scattering in SANC",
hep-ph/0611188

\bibitem{Vegas}
G.P.~Lepage,
J. Comput. Phys. 27 (1978), p.192.


\end{thebibliography}\endgroup
\end{document}